\documentclass[twocolumn]{aastex6}

\usepackage{apjfonts}
\hypersetup{citecolor=cyan}
\slugcomment{current draft: \today}

\shorttitle{Multiwavelength observations of A0620-00}
\shortauthors{Din\c{c}er et al.}

\begin{document}
\title{Multiwavength Observations of the Black Hole X-ray Binary A0620-00 in Quiescence}

\author{Tolga Din\c{c}er \altaffilmark{1}, Charles D. Bailyn\altaffilmark{1}, James C.A. Miller-Jones\altaffilmark{2}, Michelle Buxton\altaffilmark{1}, Rachel K. D. MacDonald\altaffilmark{1}}
\email{tolga.dincer@yale.edu}
\altaffiltext{1}{Department of Astronomy, Yale University, P.O. Box 208101, New Haven, CT 06520-8101, USA}
\altaffiltext{2}{International Centre for Radio Astronomy Research, Curtin University, G.P.O. Box U1987, Perth, WA, 6845, Australia}

\begin{abstract}
We present results from simultaneous multiwavelength X-ray, radio, and optical/near-infrared observations of the quiescent black hole X-ray binary A0620-00 performed in 2013 December. We find that the Chandra flux has brightened by a factor of 2 since 2005, and by a factor of 7 since 2000.  The spectrum has not changed significantly over this time, being consistent with a power law of $\Gamma = 2.07\pm 0.13$ and a hydrogen column of $N_H=3.0 \pm 0.5\times 10^{21}\rm{cm}^{-2}$.  Very Large Array observations of A0620-00 at three frequencies, over the interval of 5.25--22.0 GHz, have provided us with the first broadband radio spectrum of a quiescent stellar mass black hole system at X-ray luminosities as low as $10^{-8}$ times the Eddington luminosity. Compared to previous observations, the source has moved to lower radio and higher X-ray luminosity, shifting it perpendicular to the standard track of the radio/X-ray correlation for X-ray binaries. The radio spectrum is inverted with a spectral index $\alpha = 0.74 \pm 0.19$ ($S_{\nu} \propto \nu^{\alpha}$). This suggests that the peak of the spectral energy distribution is likely to be between $10^{12}$ and $10^{14}$ Hz, and that the near IR and optical flux contain significant contributions from the star, the accretion flow, and from the outflow. Decomposing these components may be difficult, but holds the promise of revealing the interplay between accretion and jet in low luminosity systems.
\end{abstract}


\section{Introduction}\label{sect:intro}
Accreting stellar mass black holes undergo month-year long outbursts during which their X-ray luminosity reaches close to the Eddington limit ($L_{Edd}$), but spend most of their time in quiescence \citep{Dunn10, Reynolds13, Corral-Santana16, Tetarenko16}. During the outburst/quiescence cycle, the spectral characteristics of the accretion luminosity change dramatically, along with the change in the luminosity itself \citep{Remillard06}. At higher accretion rates, sources often show a thermal X-ray spectrum associated with an optically thick, geometrically thin accretion disk, whereas at lower accretion rates a hard non-thermal X-ray emission is seen in combination with radio and infrared emission associated with a compact jet. The inner regions of the accretion disk at lower accretion rates are believed to take some form of a hot, geometrically thick, radiatively inefficient accretion flow, which are able to develop jets that may make a substantial contribution to X-rays \citep[see][and references therein]{Yuan14}. These various accretion states, and the changes between them, are subjects of considerable current research \citep[e.g.][]{Fender04, Dincer12, Dincer14, Kalemci13, Kalemci16, MillerJones12, RussellTD14}. In particular, there appears to be a strong correlation between the radio emission associated with the jet and the X-ray power law emission in the hard state \citep{Hannikainen00, Corbel00, Gallo03, Coriat11, Corbel13, Gallo12, Gallo14}. The connection between these emission components is important to understand, as it represents the connection between the relativistic jet and the accretion flow that presumably powers it \citep[see][and the references therein]{Fender16}.

The quiescent state in these sources is less well understood than brighter accretion states. The term was originally loosely used to describe any situation in which an outburst does not appear to be in progress \citep{vanParadijs84, Tanaka96}. In the X-ray band, the flux was often too low to be observed, while, in the optical, the light was dominated by the companion star, so the radial velocity curves could be observed and the mass function determined \citep{Kreidberg12, Casares14}. Subsequently, X-ray sensitivities improved and quiescent fluxes could be measured for many systems, with a range of several orders of magnitude from $10^{30.5-33.5}$ erg s$^{-1}$ \citep{Remillard06}, generally correlated with the orbital periods down to $\gtrsim 3$ hr, in the sense that longer orbital periods had higher quiescent luminosities \citep{Menou99, Garcia01, Gallo08, Homan13}. In the optical, contributions from the accretion flow were also observed, which complicated measurements of the ellipsoidal variations and thus the determination of the orbital inclination \citep{Cantrell10, Kreidberg12}. Recently, \cite{Plotkin13} have proposed a definition of quiescence as $L_X/L_{Edd} \le 10^{-5}$, this being the luminosity at which the X-ray spectral softening with decreasing luminosity seen in the hard state apparently saturates at a photon power law index of $\Gamma = 2.1$. While some change in the nature of the X-ray emission appears to occur at this luminosity, the radio/X-ray correlation seems to continue unbroken \citep{Plotkin17}. Several recent works suggested that the quiescent state may differ from the hard state having outflows with weaker particle acceleration \citep{Gallo07, Plotkin15}. Recently, there has been mounting evidence for significant variability within quiescence \citep[e.g.][]{Miller-Jones08, Hynes09, Froning11, Bernardini14, Wu16}. This is not surprising, as these sources are highly variable in all other states, and at the very least the system must evolve toward the next outburst trigger. Thus it is important that observations in different wavelength regimes be obtained simultaneously, if the overall spectral energy distribution (SED) is to be considered.

A0620-00 is the prototype of this class of X-ray transients. It underwent an intense high energy outburst in 1975 \citep{Elvis75}, and has been in quiescence ever since.  A previous outburst has been identified from archival optical plates in 1917 \citep{Eachus76}. A0620-00 was the first X-ray binary to have a mass function measured at greater than 3 solar masses \citep{McClintock86, Neilsen08} and thus was the first "dynamically confirmed" black hole candidate. Later its black hole mass was estimated to be $\rm M_{BH} = 6.6 1\pm 0.25$ $M_{\odot}$ \citep{Cantrell10}. In quiescence, A0620-00 is one of the faintest of the X-ray transients, with luminosity $\lesssim10^{-8} L_{Edd}$. But because it is relatively nearby, located at a distance of $\rm 1.06 \pm 0.12~kpc$ \citep{Cantrell10}, the quiescent emission can still be studied. Thus A0620-00 represents an opportunity to study the accretion state and the disk--jet connection at the lowest possible accretion rates. A previous multiwavelength study by \cite{Gallo06} showed that the radio/X-ray correlation established in the hard state continued down to these very faint luminosities. 

Here we present new observations of A0620-00 in quiescence, including radio, optical/near-infrared (O/NIR), and X-ray data. These observations have two features of note: (1) they are strictly simultaneous, performed within a period of 9 hr (and thus within $\approx 1$ orbital period), and thus any ambiguity relating to the known long-term variability of the source is removed; (2) we have high-frequency radio observations, which allow us to directly measure the spectral slope in the radio regime. The change in the radio/X-ray luminosity ratio is also of interest. This paper is structured as follows: Section~\ref{sect:obs} describes in detail the X-ray/radio/O/NIR observations, Section~\ref{sect:res} presents additional analysis and results, and Section~\ref{sect:disc} discusses the implications of the results.

\section{Observations}\label{sect:obs}
\subsection{Chandra}\label{sect:obs-xray}
The X-ray data were taken with the ACIS instrument on board Chandra on 2013 December 9th, starting at 00:52 UT, for $\sim$30 ks (MJD $56635.21\pm0.17$, [ObsId 14656; PI: Buxton]). The initial data processing was done following the standard threads on the Chandra X-ray Center website\footnote{\href{http://cxc.harvard.edu/ciao/threads/}{http://cxc.harvard.edu/ciao/threads/}} and using the CIAO v4.8 tools \citep{Fruscione06} and CALDB v2.7.0. No obvious flaring events were detected in the background light curve, so all the data were deemed useful.

Since A0620-00 is known to vary over time, we wanted to determine whether this most recent X-ray observation differed significantly from the previous ones. Therefore, we additionally analyzed the two archival Chandra observations of the source, performed in 2000 February \citep[ObsId 95,][]{Kong02} and 2005 August \citep[ObsId 5479,][]{Gallo06}. These older observations were reprocessed using the same calibration information, software versions, tasks, and parameters as our most recent observation.

A0620-00 was clearly detected at R.A = $06^h22^m44^s.54$, Dec $= -00^{\circ}20\arcmin44\arcsec.48$ (equinox J2000.0) in all three data sets, which is in good agreement with its optical position \citep{Liu07}. For spectral analysis, we extracted photons from a circular region with a radius of 4$\arcsec$ and a background spectrum from an annulus with an inner radius of 15$\arcsec$ and an outer radius of 25$\arcsec$, centered on the source. 

\subsection{Very Large Array}\label{sect:obs-radio}
Radio observations were performed with the Karl G. Jansky Very Large Array (VLA) on 2013 December 9th, from 03:28 to 09:27 UT (MJD $56635.28\pm0.12$), under program code SE0720 (PI: Buxton). The observing time was split between the 4--8 and 18--26\,GHz basebands (C and K bands), achieving 86 and 58 minutes on target, respectively.  In the 4--8\,GHz band, we observed in two separate 1024 MHz basebands, centered at 5.25 and 7.45\, GHz.  Each baseband was made up of eight 128 MHz spectral windows, each comprised of sixty-four 2 MHz channels.  In the 18--26\,GHz band, we observed in 3-bit mode, using four 2048 MHz basebands to cover the entire 8\,GHz of available bandwidth. Our integration time was 3\,s, and the array was in the moderately extended B-configuration.

\begin{figure}
\epsscale{1.15}
\plotone{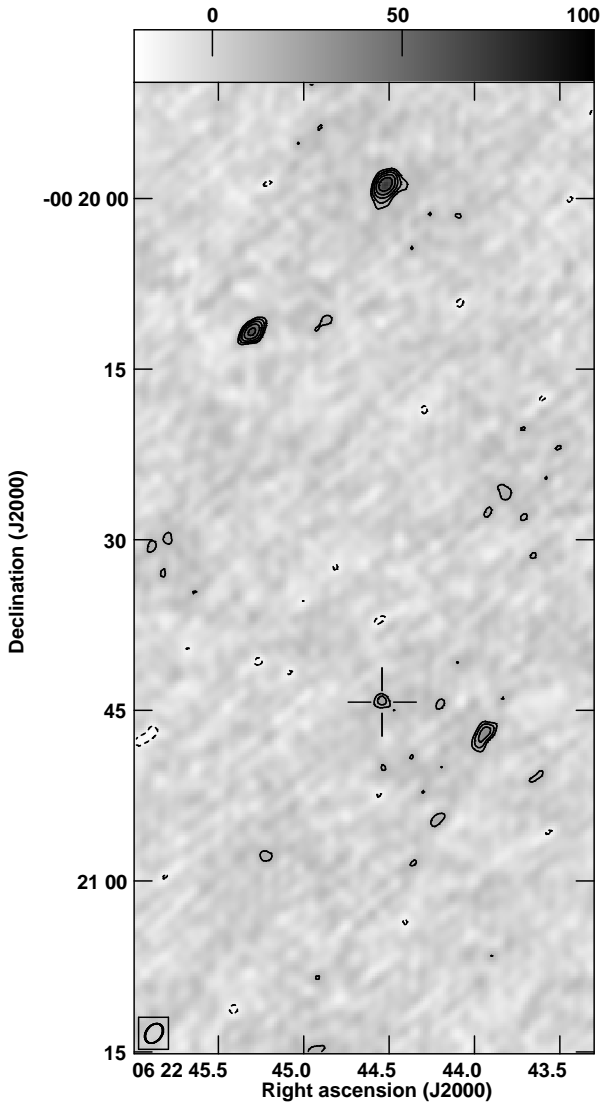}
\caption{5.25 GHz VLA radio image of A0620-00 and its surroundings, covering the same field of view as figure 2 of \citet{Gallo06}.  Contours are at $\pm(\sqrt{2}^n)$ times the rms noise level of 4.8\,$\mu$Jy\,beam$^{-1}$, with $n=3,4,5,...$.  The beam size is $1.89\times1.35$\,arcsec$^2$ at PA $-37.6^{\circ}$, and the gray scale runs from $-20$ to 100\,$\mu$Jy\,beam$^{-1}$.  A0620-00 (marked with a cross) is clearly detected, as are the N and NE sources mentioned by \citet{Gallo06}.}
\label{fig:radiomap}
\end{figure}

In both frequency bands, we used 3C147 to set the amplitude scale and perform both instrumental delay and bandpass calibration, and the nearby, compact calibrator source J0641-0320 (5$^{\circ}$.6 away) to determine the time-varying complex gain solutions. Data were reduced according to standard procedures within the Common Astronomy Software Application \citep[CASA;][]{McMullin07}, v4.2.0. Following external gain calibration, the target data were imaged separately at 5.25, 7.45, and 22\,GHz. Since the field contained no bright, confusing sources above 0.2\,mJy\,beam$^{-1}$, we created naturally weighted images to maximize our sensitivity. An example radio map at 5.25 GHz is shown in Figure~\ref{fig:radiomap}. A0620-00 was detected in all three frequency bands, and in no case was there sufficient emission in the field to attempt self-calibration. We measured the target flux density by fitting a point source in the image plane, using the CASA task {\sc imfit}.

\subsection{Optical and Near-infrared}\label{sect:obs-oir}
O/NIR observations of A0620-00 were carried out at Cerro Tololo Inter-American Observatory (CTIO) with the dual-channel imager ANDICAM \citep{Depoy03} on the SMARTS 1.3m telescope \citep{Subasavage10} using B+K, V+J, I+H filters. The observations were taken between 2013 December 6 03:28 UT (MJD 56632.14) and 2013 December 12 08:04 UT (MJD 56638.34), covering a period of 7 days around the night of the simultaneous Chandra/VLA observations. 

We reduced the O/NIR data in IRAF \citep{Tody86, Tody93} following the standard procedures described in \cite{Buxton12}. After eliminating problematic images by visual inspection, we retained 276 images in our analysis, 116 of which belong to the night of simultaneous Chandra/VLA observations. The optical images had exposure times of 300 s in B and 240 s in I and V, while the NIR images consisted of six dithered images in H and J, and seven dithered images in K, with each dithered image having a 30 s exposure time in all bands. We performed point-spread function photometry to measure the instrumental magnitudes from the source and several nearby field stars, and then converted the instrumental magnitudes of the source to the standard photometric system through the differential photometry technique with respect to nearby stars in the field, with absolute calibration via optical primary standards \citep{Landolt92} on clear nights and the Two Mass All-Sky Survey catalog \citep{Skrutskie06}. For calculating the intrinsic source fluxes, we used the zero points given in \cite{Bessell98} and the color excess $E(B-V) = 0.30 \pm 0.05$ \citep{Cantrell10}, which was converted to total extinction values $\rm A_B = 1.23$, $\rm A_V = 0.93$, $\rm A_I = 0.47$, $\rm A_J = 0.26$, $\rm A_H = 0.16$, and $\rm A_K = 0.11$ \citep{Fitzpatrick99}.

\section{Analysis and Results}\label{sect:res}
\subsection{X-Ray Spectra}\label{sect:Xray}
We performed X-ray spectral analysis with the Sherpa fitting package \citep{Freeman01}. As can be seen from Table-\ref{table:Chandra}, the net source counts are relatively low in all observations. Therefore, we grouped the spectra to have at least one photon per spectral bin; discarded the photons below 0.5 keV or above 7.0 keV; and adopted the Cash statistic \citep{Cash79} as the minimization method. Reported uncertainties (1$\sigma$) correspond to changes in the $C$ statistic of $\rm \Delta C= 1.0$ for one parameter of interest.

\begin{figure}
\epsscale{1.2}
\plotone{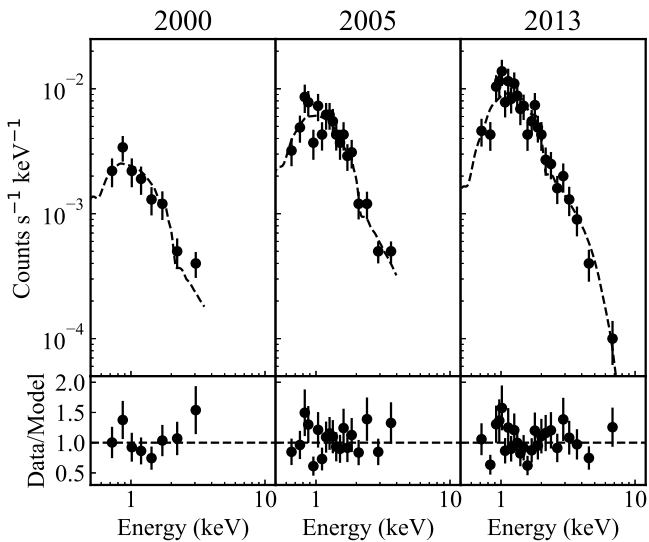}
\caption{Chandra X-ray spectra of A0620-00 from the years 2000, 2005, and 2013 with best-fit absorbed power law models and data/model ratio.}
\label{fig:xspectra}
\end{figure}

We used an absorbed power law ($\rm tbabs \times powerlaw$) with the ISM abundances of  \cite{Wilms00} and the cross sections of \cite{Verner96} to describe the X-ray spectra. We began fitting each spectrum individually. The best-fit column density and photon index values were, within large error bars, consistent with each other. In order to obtain the best constraints on the fit parameters, we decided to fit the three data sets simultaneously and tied the $N_H$ between the observations. This combination of models fitted the data well (Goodness\footnote{Goodness is the fraction (\%) of fits to the simulated spectra, which yield a lower C-stat value than the best fit to the real data.} = 0.03\%, $C-stat / dof$ = 717/851. The best-fit models on top of binned X-ray spectra are shown in Figure~\ref{fig:xspectra}. We estimated the unabsorbed source flux from each observation using the best-fitting power laws of the joint fit and then calculated the corresponding Eddington-scaled luminosities using the distance and mass estimations given in \cite{Cantrell10}. 

The resulting fit parameters, fluxes, and the corresponding luminosities are tabulated in Table~\ref{table:Chandra}. Our new observation is brighter by a factor of $\sim$2 than the 2005 observation and by a factor of $\sim$7 than the 2000 observation at $L_X \sim 10^{-8} L_{Edd}$ (see also Figure~\ref{fig:xspectra}). Although the X-ray flux increased, this change does not seem to have affected the photon index. The photon index $\Gamma = 2.07 \pm 0.13$ is, within error bars, constant over 13 years. The column density from the combined fit is $N_H = (3.0 \pm 0.5) \times 10^{21}$ cm$^{-2}$. Note that this is the tightest constraint ever obtained from X-ray observations for the $N_H$ toward A0620-00. Using the transformation law in \cite{Foight16}, it corresponds to a color excess of $E(B-V)=0.34 \pm 0.06$. \cite{Wu83} estimated the color excess toward A0620-00 to be $E(B-V)=0.35 \pm 0.02$ from the localized extinction curves. \cite{Cantrell10} found a color excess of $E(B-V)=0.30 \pm 0.05$ by modeling many different long-term quiescent O/NIR light curves of A0620-00. Our color excess is, within error bars, consistent with these commonly used values.

\begin{deluxetable*}{cccccccccc}
\tabletypesize{\scriptsize}
\tablecolumns{10} 
\tablecaption{X-Ray Spectral Fit Results \label{table:Chandra}}
\tablehead{\colhead{ObsId} & \colhead{MJD} & \colhead{Time on Source} & \colhead{Net Count Rate} & \colhead{Net Source Counts} & \colhead{Background Counts}& \colhead{$N_H$} & \colhead{$\Gamma$} & \colhead{F$_{3-9keV}$} & \colhead{L$_{3-9keV}/L_{Edd}$}\\
\colhead{} & \colhead{(day, UTC)} & \colhead{(ks)} & \colhead{0.5--7 keV (counts s$^{-1}$)} & \colhead{0.5--7 keV} & \colhead{0.5--7 keV} & \colhead{($10^{21}$ cm$^{-2}$)} & \colhead{} & \colhead{($10^{-14}$ cgs)} & \colhead{($10^{-9}$)}\\
\colhead{(1)} & \colhead{(2)} & \colhead{(3)} & \colhead{(4)} & \colhead{(5)} & \colhead{(6)} & \colhead{(7)} & \colhead{(8)} & \colhead{(9)} & \colhead{(10)}}
\startdata
95 & 51603.15 & 42.1 & (3.22 $\pm$ 0.28) $\times$ 10$^{-3}$ & 136 & 104 & 3.0 $\pm$ 0.5 & 2.17 $\pm$ 0.20 & $0.9^{+0.4}_{-0.2}$ & $1.5^{+0.7}_{-0.5}$ \\[0.7ex]
5479 & 53602.36 & 39.6 & (8.06 $\pm$ 0.45) $\times$ 10$^{-3}$ & 320 & 59 & 3.0 $\pm$ 0.5 & 2.32 $\pm$ 0.16 & $2.2^{+0.7}_{-0.5}$ & $3.6^{+1.4}_{-1.2}$ \\[0.7ex]
14656 & 56635.21 & 29.7 & (1.39 $\pm$ 0.07) $\times$ 10$^{-2}$ & 413 & 53 & 3.0 $\pm$ 0.5 & 2.07 $\pm$ 0.13 & $6.1^{+1.7}_{-1.3}$ & $9.9^{+3.7}_{-3.2}$
\enddata
\tablecomments{Column (1): Observation Id. Column (2): Modified Julian date (MJD = JD-2400000.5). Column (3): Exposure time of the observation. Column (4): Total count rate in the source aperture after background subtraction. Column (5): Number of counts in the source aperture after background subtraction. Column (6): Number of background counts in the source aperture. (7): Hydrogen column density ($N_H$), tied between the observations. Column (8): Photon index of the power law. Column (9): Unabsorbed net source flux in the $3-9$ keV band. Column (10): Luminosity in the $3-9$ keV band scaled in Eddington units.  Note that the fraction of the total luminosity implied by the observed $3-9$ keV band depends on assumptions about the shape of the overall spectrum, and thus $L_{\rm bol}/L_{\rm Edd}$ will be larger than the value given in this column. Errors on the fit parameters refer to the 1$\sigma$ uncertainties. Errors on the luminosities include uncertainties in flux, mass, and distance.}
\tabletypesize{\small}
\end{deluxetable*}

\subsection{Radio Spectrum and Radio/X-Ray Correlation}\label{sec:radio}
The measured radio flux densities at three frequency bands are given in Table\,\ref{tab:VLA}. These radio flux densities correspond to a power law of spectral index $\alpha$ = 0.74 $\pm$ 0.19 ($\alpha$ is defined according to $S_{\nu} \propto \nu^{\alpha}$ throughout the paper, where $S_{\nu}$ is the flux density at frequency $\nu$).

\begin{figure}
\epsscale{1.26}
\plotone{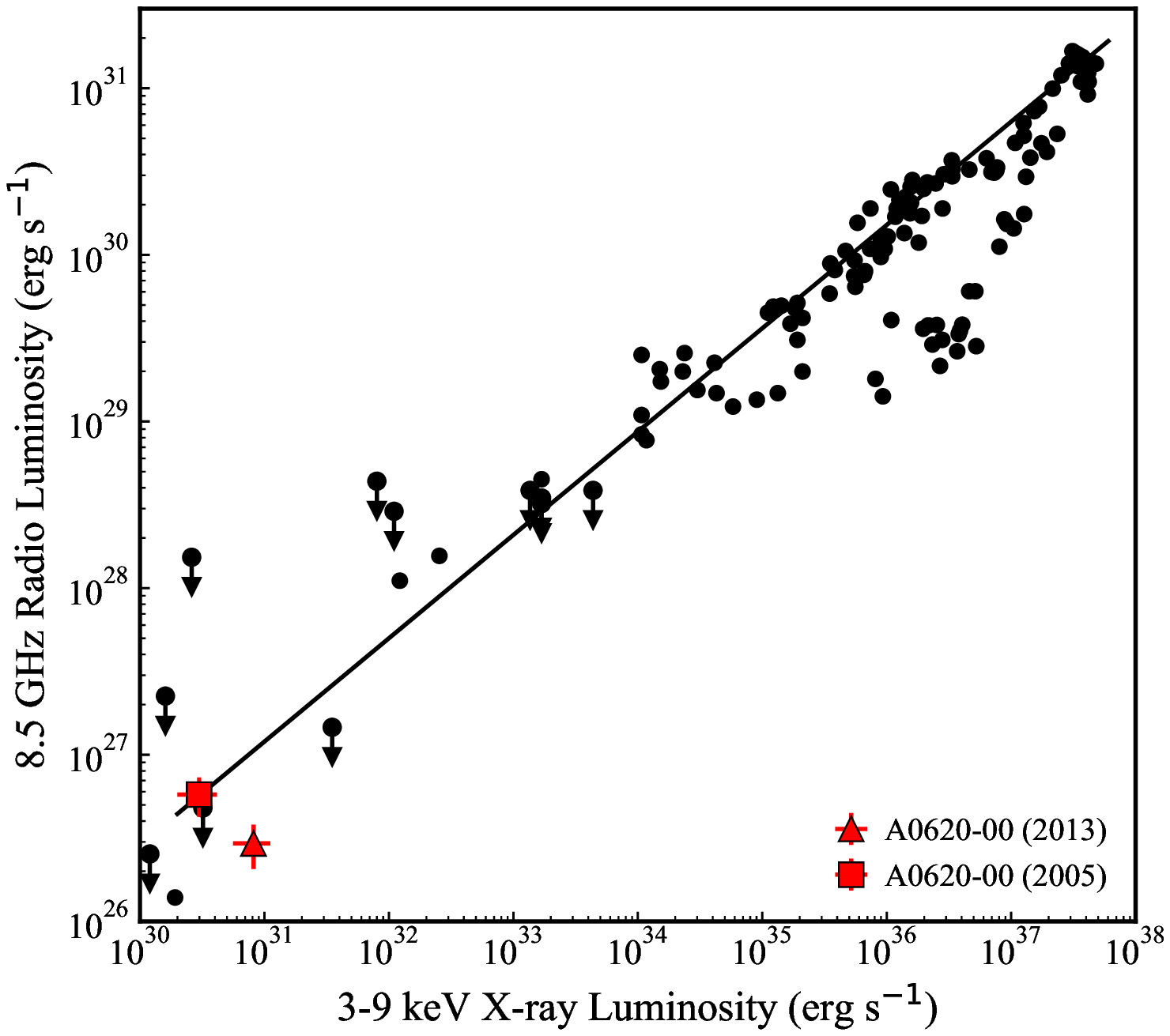}
\caption{Radio and X-ray luminosities of black hole X-ray binaries in the hard and quiescent state, highlighting A0620-00 observations in 2013 December (triangle) and 2005 August (square). Data points for other black holes include GX 339-4 \citep{Corbel13}, V404 Cyg \citep{Corbel08, Rana16, Plotkin17}, H1743-322 \citep{Coriat11}, XTE J1118+480 \citep{Gallo14}, and some quiescent sources from a deep radio survey \citep{MillerJones11}. The solid line is drawn to indicate the standard track ($L_R \propto L_X^{0.62}$, \citealt{Corbel13}). All radio luminosities are calculated by assuming a flat radio spectrum up to 8.5 GHz (i.e., $\rm L_R = 4\pi d^2 \nu S_{\nu}$). The radio luminosity of the 2013 data is calculated using the interpolated flux density at 8.5 GHz (see Section~\ref{sec:radio}).}
\label{fig:radioX-ray}
\end{figure}

To make an accurate comparison with the 2005 radio observation, we interpolated the best-fitting radio power law  and adopted a conservative 30\% fractional error, similar to what is observed in other bands. We find a resulting flux density of 26 $\pm$ 8 $\mu$Jy at 8.5 GHz. In contrast to our expectation from the increase in the X-rays, the radio flux density has dropped by a factor of two since 2005. When we add our data point to the radio/X-ray luminosity plane, we see that the 2013 data point is a little below but within the scatter of the extension of the standard track to the quiescent state, as shown in Figure~\ref{fig:radioX-ray}.

\begin{deluxetable}{cccc}
\tablecolumns{3}
\tablecaption{Measured radio flux densities of A0620-00 \label{tab:VLA}}
\tablewidth{\columnwidth}
\tablehead{\colhead{Central Frequency} & \colhead{Bandwidth} & \colhead{Flux Density} \\
\colhead{(GHz)} & \colhead{(GHz)} & \colhead{($\mu$Jy\,beam$^{-1}$)}}
\startdata
5.25 & 1 & $22.3\pm4.8$\\
7.45 & 1 & $18.9\pm5.3$\\ 
22.00 & 8 & $53.6\pm5.0$
\enddata
\end{deluxetable}

\subsection{Limits on the Intra-night Radio and X-Ray Variability}\label{sec:intravar}
We searched for intra-night variability in our VLA and Chandra observations. Examining the 22 GHz data, we find marginal evidence for a change in radio flux. The 22 GHz data was obtained in two observing windows, from 04:27 to 06:18 UT and 07:00 to 08:50 UT. Splitting each of the observing windows into two equal time bins, we detected the source at 50 $\pm$ 10 $\mu$Jy beam$^{-1}$ and  44 $\pm$ 11 $\mu$Jy beam$^{-1}$ during the first observing window, and 55 $\pm$ 9 $\mu$Jy beam$^{-1}$ and 80 $\pm$ 10 $\mu$Jy beam$^{-1}$ in the second observing window. Given the large error bars, a constant source cannot be ruled out at the 2$\sigma$ level, but it is probable that the radio flux varies on a timescale of hours - shorter timescales cannot be explored given the low flux levels. The other two radio bands have $<5\sigma$ detections for the entire observation, so subdividing them in time yields no useful information. In order to investigate the X-ray variability during the radio observations, we extracted a background-subtracted source light curve in the 0.5-7 keV band with time bins of 3600 s. We found that the errors on the count rates are greater than the bin-to-bin variability. The fractional errors on the count rates are 15\%, hence we are not able to detect any variability smaller than 15\% on hour timescales.

\subsection{O/NIR Fluxes}\label{sec:onirfluxes}
O/NIR magnitudes of A0620-00 were significantly variable in all our observing bands on the night of our Chandra/VLA observations. Comparing the ranges of VIH magnitudes to Fig~2 in \cite{Cantrell08}, we identify A0620-00 as being in the active state, where the O/NIR emission shows enhanced non-stellar emission.  The O/NIR light is not straightforward to interpret, as it contains emission from several different sources in the system.  In quiescent systems such as this, the companion star provides a substantial fraction of the O/NIR but there are also contributions from the accretion flow, and possibly from the jet as well. In the remainder of this section we consider how to separate these components. 

\subsubsection{Stellar and Non-stellar Light}
The problem of separating the companion star from the other light in the system has been addressed in the context of measuring the ellipsoidal variations generated by the changes in the geometric cross-section of the companion star with orbital phase. In the case of A0620-00, this has been done in some detail by \cite{Cantrell10}. They used many different orbital light curves to determine a consistent model for the orbital elements and the companion star, which are not expected to change with time, together with a variable component associated with the accretion flow. We followed this approach by removing the stellar flux from the total flux in all observing bands. In quiescence, this removal process had to take into account the phase-dependent change of the stellar emission. 

\begin{figure*}
\epsscale{1.25}
\plotone{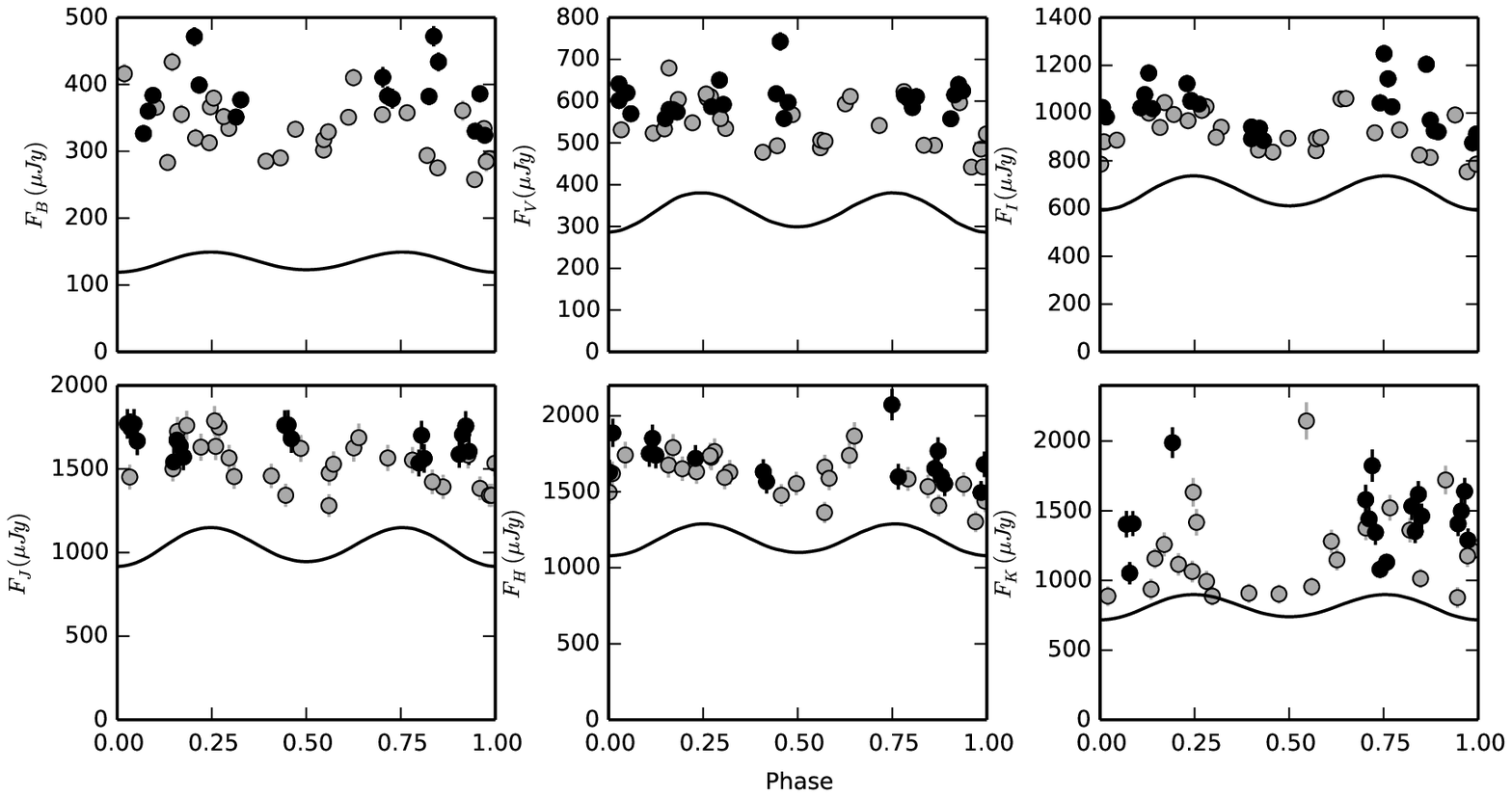}
\plotone{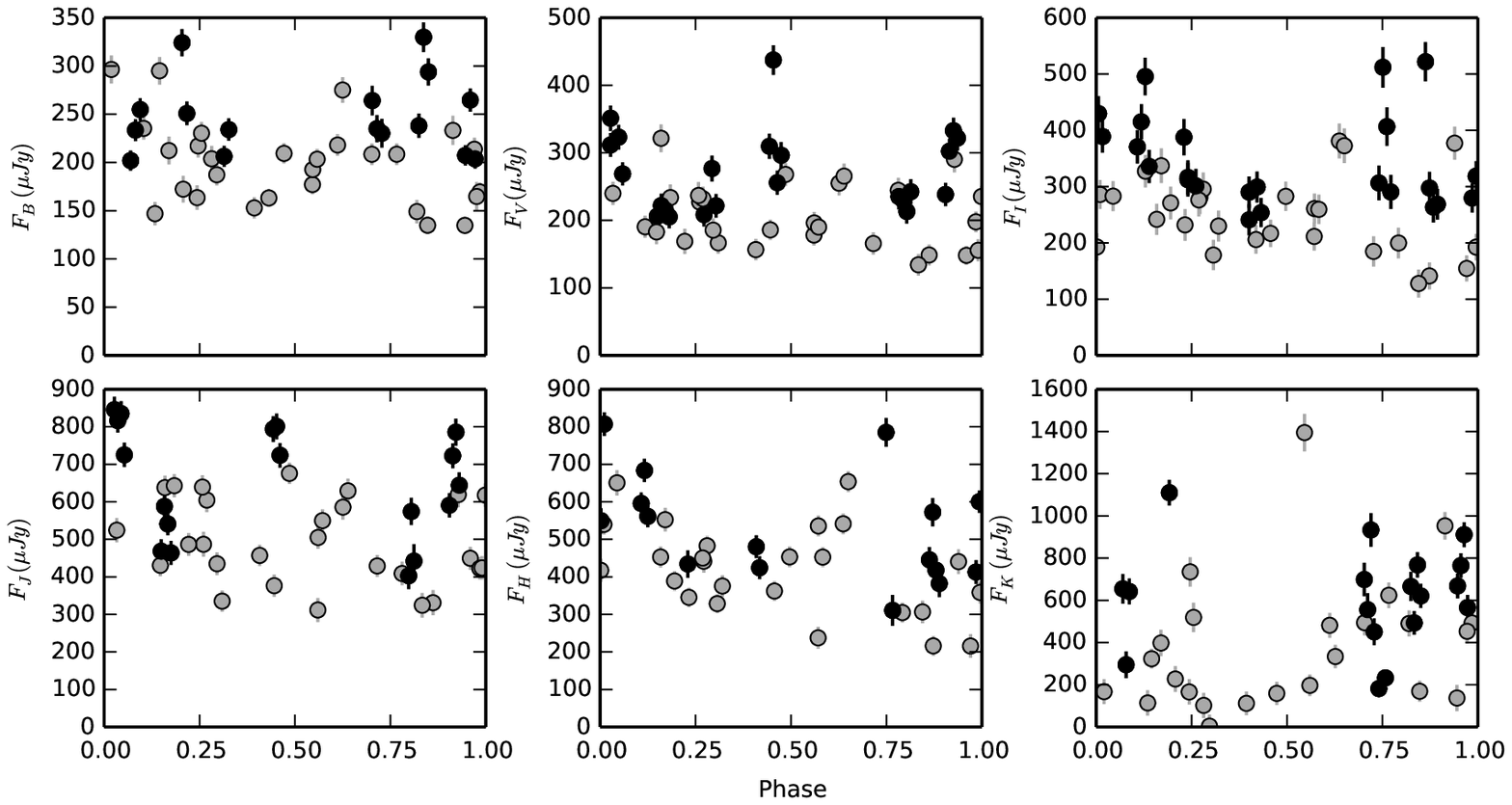}
\caption{Phase-folded O/NIR light curves of A0620-00, covering a time period between 2013 December 6 03:28 UT (MJD 56632.14) and 2013 December 12 08:04 UT (MJD 56638.34). Black circles show the observations taken on the night of 2013 December 09/10 whereas gray circles show the data from other dates. Top six panels show the total emission along with model stellar emission (solid lines). Bottom six panels show the non-stellar emission. All fluxes are de-reddened. Error bars contain only the photometric uncertainties, the systematic error due to absorption correction is not included.}
\label{fig:optirlc}
\end{figure*}

We made use of the VIH-band stellar light curves of A0620-00 presented in \cite{Cantrell10}. First, we folded our light curves on the same period ($\rm P= 0.323014~d$) and ephemeris ($\rm T_0 = 2454084.85635~JD$) as that used by \cite{Cantrell10} and then subtracted the stellar fluxes from our measurements. The stellar flux in the BJK bands was not derived in \cite{Cantrell10}. For these bands, first we interpolated the model stellar light curve in the VIH bands to the BJK bands using the intrinsic colors of a K5V star \citep{Bessell88} and then subtracted these transformed fluxes from the total BJK fluxes. The phase-dependent evolution of the total and non-stellar fluxes, and the stellar model fluxes for all bands are shown in Figure~\ref{fig:optirlc}.

Next we calculated the average total and non-stellar fluxes, and their fractional variability amplitudes for each band. The fractional variability F$_{var}$ and its uncertainty $\sigma_{F_{var}}$ are defined as 

\begin{equation}
F_{var} = \sqrt{\frac{S^2-\overline{\sigma^2_{err}}}{\overline{x}^2}}
\end{equation}

\begin{equation}
\sigma_{F_{var}} = \sqrt{\left(\frac{1}{2N}\frac{\overline{\sigma^2_{err}}}{\overline{x}^2F_{var}}\right)^2+\left(\sqrt{\frac{\overline{\sigma_{err}^2}}{N}}\frac{1}{\overline{x}}\right)^2}
\end{equation}

where $\overline{x}$ is the average flux, $S$ is the standard deviation of the $N$ flux measurements, and $\overline{\sigma_{err}^2}$ is the mean of squared error of those $N$ measurements, all for a given band \citep{Edelson02, Vaughan03}. The mean de-reddened total and non-stellar fluxes obtained from the night of simultaneous Chandra/VLA observations along with their associated fractional variabilities are given in Table~\ref{tab:oirflux}. Note that the variability of the non-stellar emission increases with wavelength, with $F_{var}$ = 15.7 $\pm$ 0.3 \% for $B$ and $F_{var}$ = 41.6 $\pm$ 0.6 \% for $K$ whereas the variability of the total emission is lower and shows no clear trend with wavelength.

\subsubsection{Decomposing the Non-stellar Emission}
As noted above, the non-stellar light may itself contain multiple components, possibly including contributions from a symmetric accretion disk, one or more hotspots, and the jet.  On the timescales of our observations ($\approx 1$ orbital period) one expects the accretion disk to be stable, and the hot spot light to vary with orbital phase, but light from the jet may be variable on much shorter timescales.  Short timescale variability superimposed on ellipsoidal variation in quiescence has been observed in detail from this source and several other soft X-ray transients \citep{Hynes03, Zurita03, Shahbaz04}. We have therefore examined the variability characteristics of the non-stellar O/IR light in an effort to provide additional constraints on its emission components.

We find that there is considerable short-term variability left over after the companion star has been removed (see Figure~\ref{fig:optirlc}).  We then attempted to decompose the non-stellar flux into a non-varying continuum flux and variable flux on top of this continuum.  Given the relatively few orbits observed, it is hard to distinguish orbital variability from random flares, and there is no compelling evidence in Figure~\ref{fig:optirlc} for orbital variability as such, although it is not ruled out.  There does appear to be a floor level of non-stellar light, with flaring above that. We attempted to determine the lower floor in each bandpass as follows: starting with the non-stellar data from all days, we iteratively exclude the data that is more than 1$\sigma$ greater than the mean flux until no data is left to be excluded.  After that we averaged the remaining measurements, which are typically around 10 in number in each bandpass. The fluxes of the continuum and flare components determined in this manner are given in Table~\ref{tab:oirflux}.  The relative strength of the variable component compared to the continuous flux appears to rise as one progresses to the red, as might be expected if the emission comes from a combination of a relatively blue thermal disk and redder emission from a variable jet.

\begin{deluxetable*}{ccccccc}
\tablecolumns{7}
\tablecaption{Time Averaged O/NIR Flux and Variability Results. \label{tab:oirflux}}
\tablehead{\colhead{Filter} & \colhead{Total Flux} &  \colhead{$F_{var}^{Tot}$} &  \colhead{Non-Stellar Flux}  & \colhead{$F_{var}^{NS}$} & \colhead{Continuum Flux} &  \colhead{Flaring Flux} \\
\colhead{} & \colhead{($\mu$Jy)}  & \colhead{(\%)} & \colhead{($\mu$Jy)} & \colhead{(\%)} & \colhead{($\mu$Jy)} & \colhead{($\mu$Jy)}}
\startdata
B  & 381 $\pm$ 44 & 11.1 $\pm$ 0.10 & 245 $\pm$ 39 & 15.7 $\pm$ 0.3 & 150 $\pm$ 13 & 95 \\	
V  & 603 $\pm$ 39 & 6.37 $\pm$ 0.03 & 268 $\pm$ 57 & 21.5 $\pm$ 0.3 & 160 $\pm$ 15 & 108  \\ 
I   & 1011 $\pm$ 100 & 9.80 $\pm$ 0.04 & 340 $\pm$ 79 & 23.5 $\pm$ 0.4 & 175 $\pm$ 28 & 165 \\
J  & 1665 $\pm$ 85 & 4.44 $\pm$ 0.05 &  532 $\pm$ 139 & 22.0 $\pm$ 0.3 & 337 $\pm$ 25 & 195  \\
H & 1689 $\pm$ 141 & 8.08 $\pm$ 0.08 & 655 $\pm$ 147 & 25.8 $\pm$ 0.4 & 223 $\pm$ 13 & 432  \\
K & 1410 $\pm$ 235 & 15.46 $\pm$ 0.32 & 559 $\pm$ 269 & 41.6 $\pm$ 0.6 & 161 $\pm$ 37 & 398
\enddata
\tablecomments{All fluxes are de-reddened. The total and non-stellar fluxes are weighted averages and their errors are the weighted standard deviations.}
\end{deluxetable*}

\subsubsection{O/NIR Spectra}\label{sec:oir}
\begin{figure}
\epsscale{1.25}
\plotone{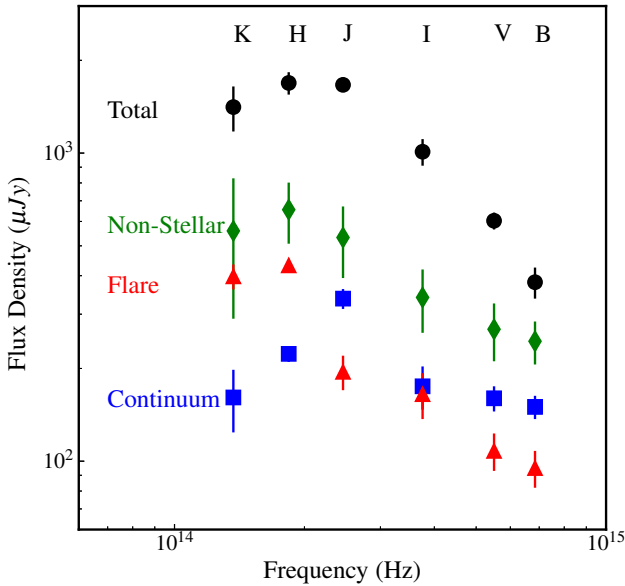}
\caption{De-reddened O/NIR spectra of A0620-00 (see Table~\ref{tab:oirflux} for the flux densities), circles showing the total flux, diamonds the non-stellar flux, triangles the flaring flux, and squares showing the continuum flux.}
\label{fig:oirsed}
\end{figure}

In Figure~\ref{fig:oirsed} we present broadband spectra for the various components we identified in the O/NIR light curves. The circles show the total light (before the stellar companion is subtracted), while the squares show the non-stellar light. The continuum and flare components of the non-stellar light are presented with diamonds and triangles, respectively.

The spectra in Figure~\ref{fig:oirsed} show distinguishing characteristics. The total light exhibits a blackbody-like shape, which peaks between $J$ and $H$ bands as expected from the bright K5V donor star in A0620-00. The non-stellar spectrum can be characterized as a power law of spectral index $\alpha = -0.68 \pm 0.08$.  While the $K$ band appears to drop, the errors are sufficiently large that there is no contradiction to a power law rising to the red. The flaring component of the non-stellar spectrum is also well characterized with a power law but with an even steeper spectral index $\alpha = -1.10 \pm 0.22$. The overall shape of the non-variable component appears to be flat, with significant excess at $J$ and $H$ bands, although this may be an artifact of our technique for separating the non-variable and flaring components.  If real, this peak may be associated with the peak predicted for emission from an ADAF component \citep[e.g.][]{Quataert99}.

\subsection{Broadband Spectrum}\label{sec:broadband}
\begin{figure*}
\epsscale{1.2}
\plotone{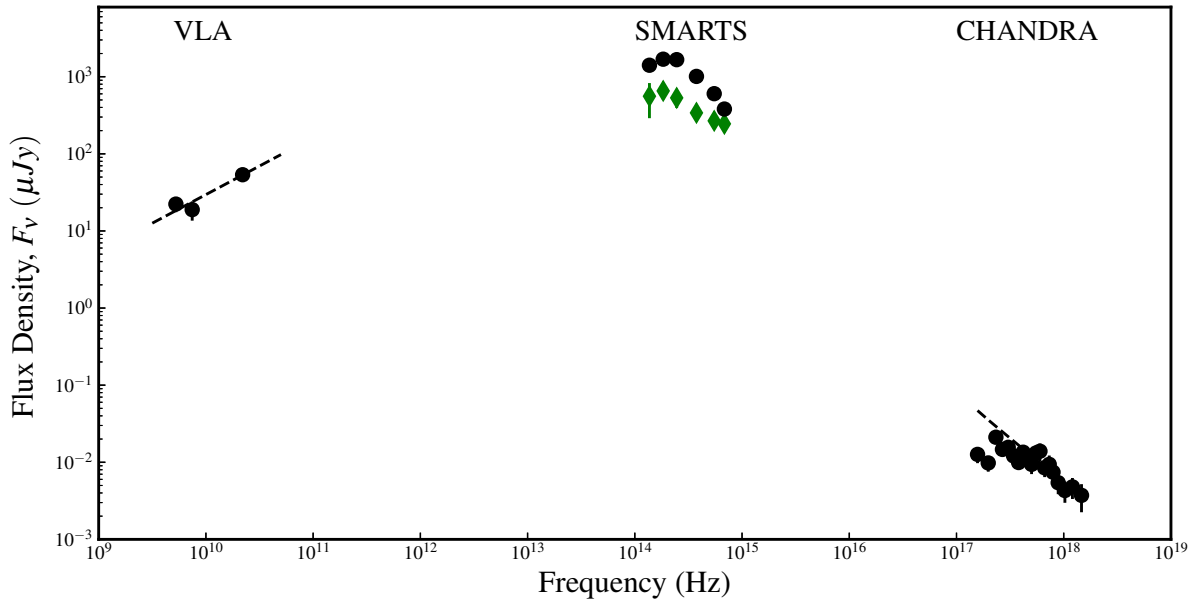}
\caption{Broadband spectrum of A0620-00 constructed from VLA, SMARTS, and Chandra observations taken on 2013 December 9. SMARTS O/NIR data are de-reddened and consist of total light (circles) and non-stellar light (diamonds). Solid lines show the best fitting power laws to radio and unabsorbed X-ray data.}
\label{fig:sed}
\end{figure*}

In the previous sections, we presented the results of data from individual wavelength regimes. Here we show the broadband spectrum of A0620-00 in Figure~\ref{fig:sed}.  Since the radio and X-ray emission are not expected to contain significant flux from the star, we focus on the connection between the non-stellar O/NIR data and the data in other wavelength regimes.  As we showed in Section~\ref{sec:radio}, the radio spectrum is a highly inverted power law. Extrapolation of this radio power law exceeds the flux measured in near-infrared. The assumption that the flux takes the form of a broken power law implies that the broadband spectrum (companion subtracted) would peak somewhere between $10^{12}$ and $2\times 10^{14}$ Hz.\   The lower break frequency implies that the O/NIR spectrum remains flat into the thermal IR, while the radio continues its inverted spectrum toward higher frequencies.  The higher break frequency implies that the O/NIR flux peaks near the K-band.
 A formal broken power law fit to the radio and non-stellar O/IR emission suggests a spectral peak at $(9 \pm 5) \times 10^{12}$ Hz, but if either component is something other than a perfectly straight power law \citep{Peer09}, the location of the break can change significantly.  Using only the flaring component in the O/NIR does not change the fitted break frequency significantly.  

When looking at higher frequencies of the spectrum, we see that the extrapolation of the non-stellar O/IR power law is well above the observed X-ray flux. This suggests that there is another spectral break between optical and X-rays.

\section{Discussion}\label{sect:disc}
A number of modeling exercises have been performed for multi-band spectra, from radio to X-ray, of A0620-00 and related objects \citep{Gallo07, Plotkin15, Connors17}. In particular, a recent study has analyzed a range of possible models for A0620-00 and Sgr A*, which has a comparably low value of $L/L_{Edd}$ \citep{Connors17}.  Generally speaking, such models include synchrotron emission from a jet that dominates the radio; an SSC component that may dominate the high energy emission; a stellar component that dominates the optical; and, in some cases, a component from the accretion flow that may contribute from optical through X-ray.  It is generally assumed that the accretion flow in quiescent sources takes the form of a particularly low hard state, with an outer disk component that may contribute thermal emission in the UV through IR, and an inner radiatively inefficient flow that may contribute at higher energies. Other components might include a circumbinary disk emitting in the IR, and UV emission from pre-shock synchrotron radiation.  The X-ray flux itself is unlikely to be direct synchrotron radiation, since that would imply a steepening of the radio/X-ray correlation due to synchrotron cooling that is not observed in quiescence.  In general, it is worth noting that the balance between the various emission components may be different in quiescence than in the hard state, as demonstrated by \cite{Plotkin15}.

It is beyond the scope of this paper to carry out such a detailed modeling exercise.  However, we comment here on three aspects of our results that may constrain or complicate such efforts, namely the strongly inverted radio slope; the radio/X-ray anticorrelation with respect to previous observations of A0620-00; and the various observed components of the O/NIR emission.

\subsection{Radio Slope}\label{sect:radiojet}
It has been suggested that the single frequency radio detections of quiescent black holes at luminosities similar to ours is an indication of the survival of hard state jets in the quiescence \citep{Gallo06, Gallo14}, but this hypothesis has not previously been confirmed with the detection of a flat or inverted radio spectrum. The inverted radio spectrum presented here supports the jet hypothesis in that it is within the expectations of jet models \citep{Hjellming88, Peer09}. That said, the measured radio slope is highly inverted compared to commonly observed values ($0.0 < \alpha$ < 0.3) and such a steep slope has only been observed in a few accreting black holes, such as GX 339-4 \citep{Corbel13}, MAXI J1836-194 \citep{RussellTD14}, and XTE J1118+480 \citep{Fender01}.  Typical synchrotron models, such as those used in \cite{Connors17}, tend to have a much shallower slope, and such a highly inverted spectrum is not fully compatible with standard jet models that invoke a conical geometry  \citep{Blandford79}. However, a rapidly flaring jet geometry could explain this inversion. In such a geometry, adiabatic expansion might make the emission at the outer parts of the jet fall faster than the emission at the regions closer to the black hole, and hence generate an inverted radio spectrum. To evaluate these or other possibilities for the highly inverted slope, it would be very useful to know whether our observations are anomalous, perhaps representing an uncommon or transient situation, or whether such a slope is typically seen in quiescent X-ray binaries.

\subsection{X-ray/Radio Anticorrelation}
We have shown that A0620-00 exhibited a significant change in its X-ray flux over 13 years. The change in the X-ray flux between the two archival Chandra observations was reported previously \citep{Gallo06, Plotkin13} and the flux from a combined $Swift$ observation in 2010 was consistent with the 2005 level \citep{Froning11}. Our results reinforce the fact that A0620-00 is variable in X-rays at the deepest quiescent levels probed to date at $L_X \sim 10^{-8} L_{Edd}$. Furthermore, we have also shown marginal evidence that the radio flux density of the source has changed with respect to 2005.

A0620-00 is now the only accreting black hole from which radio and X-ray variability has been observed at $L_X \sim 10^{-8}$ $L_{Edd}$. The closest source in luminosity that has shown variability in both regimes is V404 Cyg at $L_X \sim 10^{-6} L_{Edd}$ \citep{Bernardini14, Plotkin17}. The results from V404 Cyg and A0620-00 suggest that the X-ray (and perhaps the radio) variability is a common property of black hole X-ray transients in quiescence down to at least $L_X \sim 10^{-8} L_{Edd}$, and hence provides motivation for future variability studies at the deepest quiescent levels.

The origin of the X-ray variability can be interpreted quite differently depending on the timescale of the variability.  Given that there are only three observations, it may well be that we are seeing the results of stochastic variations in the X-ray flux.  While there is no evidence of short-term variability within our data (see Section~\ref{sec:intravar}), V404 Cyg in quiescence does show flares on day to week timescales during which the X-ray flux of the source increases by a factor of 3 \citep{Bernardini14}. This suggests that the increase in the X-ray flux might be due to random snapshots of these short-term flares and might thus be independent of changes in the outer parts of the accretion disk.  Alternatively, one might speculate that the X-ray flux is a long-term trend, which might result from a gradual buildup of the accretion disk for the next outburst cycle. This would not be unexpected. Brightening in the long-term O/NIR light curves of accreting black hole transients in quiescence has been observed in several sources \citep{Cantrell08, Wu16, Koljonen16}. The changes in the outer disk and inner flow (as observed from O/NIR and X-rays, respectively) may both be the consequence of the buildup of the disk, which is expected by the instability model that explains the outburst cycles \citep{Dubus01}.  More repeated observations during quiescence will be useful to understand the origin of the variability.

Regardless of a possible connection between the inner accretion flow and the outer accretion disk as suggested by the O/NIR and X-rays, the change in X-rays presumably reflects a change in the emission properties of the material flowing close to the black hole \citep[e.g.][]{Sobolewska11}. There is a general consensus that the X-rays are emitted by a population of electrons in this flow; however the origin and nature of these electrons are still debated. They can be either thermal or non-thermal, inflowing or outflowing \citep{McClintock03, Veledina13}. However, we find that the power law shape and index of the X-ray flux does not change as the flux increases. Therefore, it seems likely that the nature of the electron population responsible for the X-ray flux has not changed, although the size of the emitting region may have increased.

The radio emission, which we interpret as the signature of jets in quiescence, is likely to vary during our observation duration on hour timescales (see Section~\ref{sec:intravar}). On such timescales, fast ejecta could collide with previous slower ejecta, resulting in a shock, and hence cause variability. This shock phenomenon within a jet has previously been invoked to explain the radio properties of jets in general \citep{Jamil10, Malzac13}, and there are likely other ways to change the magnetic field or electron density on short timescales. But it is difficult to detect such variations in the radio flux of A0620-00, because the radio flux is very low (see Section~\ref{sec:intravar}). Thus our radio observations, and indeed any plausible future observations of systems at the faint end of the radio/X-ray correlation, must be interpreted as an average over what may be substantial variability on timescales that are short compared to the exposure time. Direct comparisons of jet activity as indicated by the radio with the state of the accretion flow as indicated by simultaneous X-ray observations should therefore be considered with caution.

Small changes in the X-ray/radio correlation have also been observed in sources such as GX 339-4 \citep{Corbel13}, and it is interesting to speculate whether an anticorrelation such as we observe might be directly associated with the increase in the X-ray flux. For example, if the accretion flow in 2013 is denser than in 2005 as suggested by the higher X-ray luminosity, it may cool down the electrons in the jet base, drop the number of electrons that will be channeled into the jet, and lead to fainter jet radio emission.  Before speculating further, it will be important to establish that this anticorrelation of the radio and X-ray fluxes is a pattern, rather than a single anomalous observation.

\subsection{Interpretation of the Broadband Spectrum}
The SED of the jet, including the peak  luminosity and turnover frequency, is crucial to understanding the physical properties associated with the jet and shock.  The SED of the accretion flow would provide constraints on a non-disk accretion, most likely in a radiatively inefficient state.  In both cases, the extremely low mass accretion rate in this system makes this effort particularly interesting.  It is expected, for example, that very low accretion rates would be associated with particularly inefficient flows, since the low density leads to particularly inefficient coulomb coupling - thus systems of this kind are thought to demonstrate particularly strong differences depending on whether the flow encounters an event horizon or a surface \citep{Narayan97}.

There are, however, serious difficulties in interpreting the broadband spectrum shown in Figure~\ref{fig:sed}. Most obviously, the low accretion rate results in low luminosities, hence in low photon counts -- however this is mitigated somewhat by the relatively small distance to the source and the sensitivity of the instruments.  More importantly, the peak of the spectrum appears to be in a currently inaccessible region of the spectrum. The radio spectrum is inverted and the non-stellar O/NIR rises toward longer wavelengths, suggesting a spectral peak between these wavelengths. Another complication is that there appear to be at least three different emission sources in the O/NIR. As described in Section~\ref{sec:onirfluxes}, there is the companion star, a continuum flux, and a flaring flux. A more comprehensive study of the timing and spectrum of the O/NIR emission will be required to separate these sources in a compelling way. But already we can see that while the star peaks in the NIR, the continuum and particularly the flaring component may rise into the thermal IR.
 
\subsubsection{Origin of the Non-stellar O/NIR emission}
The non-stellar emission can be interpreted as a single emission component or in the form that we decomposed it into a flare and a continuum components. Here we discuss the implications for the jet emission under both sets of assumptions: that all the non-stellar O/NIR flux comes from the jet, and that only the variable component comes from the jet.

The total non-stellar light is consistent with an optically thin synchrotron emission with spectral index $\alpha =-0.68 \pm 0.08$, which corresponds to a particle distribution of $p= (1-2\alpha) = 2.36 \pm 0.16$, \deleted{is} consistent with the standard particle acceleration theory \citep{Bednarz98, Kirk00} and commonly accepted values in jet models \citep{Heinz03, Merloni03, Falcke04}. No additional emission component is required, such as a pre-shock component that is often employed in some jet models \citep{Gallo07, Plotkin13}. In this case the variability in the non-stellar light would then be attributed to changes at the jet base. Such short-term variability in the near/mid infrared that is associated with the jet base has been observed in the hard state of GX 339-4 \citep{Gandhi11}, albeit at much higher X-ray luminosity.

Alternatively, if we consider only the flaring component, the spectral index of the power law is $\alpha = -1.10 \pm 0.22$ which corresponds to a particle distribution of $p=3.2 \pm 0.4$. Although this spectral index is quite steep for commonly observed values, \cite{Russell10} claimed that the jet spectral index in XTE J1550-564 evolved smoothly from $\alpha = -1.5$ when the jet was faint to $\alpha = -0.5$ when the jet was bright \citep[but see also][]{Poutanen14}. The very steep spectral index seen when the jet was faint was explained by a thermal, possibly Maxwellian distribution of electrons in a weaker jet and this might be similar for A0620-00 \citep{Plotkin15}.   

The non-flaring continuum shows a complex spectral shape, which is difficult to interpret.  This may indicate some ambiguity in the separation of the components.  In this context, it is  worth noting that in V404 Cyg, the X-ray light curve correlates with the $H_\alpha$ light curve during quiescence, suggesting an X-ray driven variability in the disk \citep{Hynes04}.

\subsubsection{Constraints on the jet physical properties}
Considering the radio and non-stellar O/NIR spectra, we have identified a range of locations for the jet spectral break between $10^{12}$ Hz and $2 \times 10^{14}$ Hz.  \cite{Russell13} suggested a spectral break at $1.3 \times 10^{14}$ Hz consistent with one end of our range. There is also a possibility that the jet spectral break may be varying within the night. A varying jet spectral break was detected in the hard state of GX 339-4 at much higher X-ray luminosity \citep{Gandhi11}.  Following the single-zone cylindrical approximation of  \cite{Gandhi11}, we can estimate the magnetic field at the jet base ($B$) and the jet base radius ($R$). Depending upon the precise break frequency, we can find values for $B$ between $6\times 10^3$ G and $2\times 10^5$ G, and for $R$ between $8 \times 10^{6}$ cm (4 $r_g$) and $5 \times 10^{8}$ cm (250 $r_g$).
Previous work suggests a compact jet base with $R$ $<$ $10 r_g$ \citep{Plotkin15, Connors17}. A wide range of jet break frequencies are allowed by our data, but only breaks at higher frequency would suggest a compact jet base. If the flux density of the lower frequency break was at a lower level, then that would also yield a more compact jet base. In this case, the SED might gradually change slope, rather than having a well-defined spectral break. Millimeter wave observations might help resolve the current ambiguity.

\section{Conclusions}
In this work, we have presented a detailed analysis of the radio, near-infrared, optical, and X-ray observations of A0620-00 during its quiescent state. We note that  all our data were taken strictly simultaneously, which eliminates any ambiguity associated with the variability of the source. Our new observations add to previous studies of the source \citep{Gallo06, Gallo07} in a number of ways. 

(1) We find that the X-ray flux is significantly brighter than it was in 2000 and 2005.  The source appears to have brightened by a factor of 7 over 13 years, a substantial fraction of the outburst cycle.

(2) We add another point to the low luminosity end of the radio/X-ray correlation plot. Interestingly, the source seems to have moved orthogonally to the overall correlation since 2005.

(3) We have measured the first radio spectral slope in a highly quiescent black hole X-ray binary at $10^{-8}$ $L_{Edd}$. Prior to our observation, the lowest Eddington-scaled X-ray luminosity for which a radio spectral slope had been measured was at $10^{-6}$ $L_{Edd}$ in V404 Cyg \citep{Corbel08, Rana16, Plotkin17}. We find that the radio spectrum of A0620-00 is highly inverted with a spectral slope of $\alpha = 0.74 \pm 0.19$. 

(4) We have subtracted the star from the O/NIR flux, and we find that the non-stellar flux has a slope of $\alpha = -0.68 \pm 0.08$, and thus the peak of the SED appears to be in the thermal IR.

(5) In an effort to explore the possibility that both accretion emission and jet emission contribute to the non-stellar O/NIR flux, we have attempted to separate a constant component from a flaring component.  The spectral index of the flaring component alone is steeper than the total non-stellar light, at $\alpha = -1.10\pm 0.22$, which complicates the interpretation of the overall SED.

(6) Using the two archival Chandra X-ray observations along with our own, we have constrained the hydrogen column density toward A0620-00 more tightly than the previous X-ray studies of the source and it is $N_H = (3.0 \pm 0.5) \times 10^{21}$ cm$^{-2}$.

Key questions remain on many aspects of the quiescent emission from this source, which is one of the faintest observed black hole accretors in Eddington units.  Additional simultaneous multiwavelength observations will help clarify which of the observed phenomena are generally present and need to be incorporated into models, and which might be less important stochastic effects.  

\acknowledgements
We are grateful for the contributions of Tom Maccarone, Mike Garcia, and Jerry Orosz, who provided crucial input in the initial design and execution of the observations.

J.C.A.M.J is the recipient of an Australian Research Council Future Fellowship (FT140101082). This work has been supported by the National Aeronautics and Space Administration through Chandra Award Number GO3-14046X issued by the Chandra X-ray Observatory Center, which is operated by the Smithsonian Astrophysical Observatory for and on behalf of the National Aeronautics Space Administration under contract NAS8-03060. The National Radio Astronomy Observatory is a facility of the National Science Foundation operated under cooperative agreement by Associated Universities, Inc.

\software{CIAO \citep[v4.8;][]{Fruscione06}, CASA \citep[v.4.2.0;][]{McMullin07}, IRAF \citep[v2.15.1a;][]{Tody86, Tody93}, Matplotlib \citep[v2.0.2;][]{Hunter07}, Sherpa \citep{Freeman01}}

\facilities{CXO (ACIS), VLA, CTIO:1.3m}

\bibliographystyle{apj}
\bibliography{apj-jour,blackholebib}

\end{document}